# Data-driven and machine-learning based prediction of wave propagation behavior in dam-break flood

Changli LI[a]

Zheng HAN[a,c*]

Yange LI[a,b]

Ming LI[a]

Weidong WANG[a,b]

[a] School of Civil Engineering, Central South University, Changsha, 410075, China.

[b] The Key Laboratory of Engineering Structures of Heavy Haul Railway, Ministry of Education, Changsha 410075, China

[c] Hunan Provincial Key Laboratory for Disaster Prevention and Mitigation of Rail Transit Engineering Structures, Changsha 410075, China

* Corresponding author: Z. Han (zheng_han@csu.edu.cn), 68 Shaoshan Road, Central South University, Changsha, Hunan, China. Tel.: +86 18874163071.

# Abstract

The computational prediction of wave propagation in dam-break floods is a long-standing problem in hydrodynamics and hydrology. Until now, conventional numerical models based on Saint-Venant equations are the dominant approaches. Here we show that a machine learning model that is well-trained on a minimal amount of data, can help predict the long-term dynamic behavior of a one-dimensional dam-break flood with satisfactory accuracy. For this purpose, we solve the Saint-Venant equations for a one-dimensional dam-break flood scenario using the Lax-Wendroff numerical scheme and train the reservoir computing echo state network (RC-ESN) with the dataset by the simulation results consisting of time-sequence flow depths. We demonstrate a good prediction ability of the RC-ESN model, which ahead predicts wave propagation behavior 286 time-steps in the dam-break flood with a root mean square error (RMSE) smaller than 0.01, outperforming the conventional long short-term memory (LSTM) model which reaches a comparable RMSE of only 81 time-steps ahead. To show the performance of the RC-ESN model, we also provide a sensitivity analysis of the prediction accuracy concerning the key parameters including training set size, reservoir size, and spectral radius. Results indicate that the RC-ESN are less dependent on the training set size, a medium reservoir size $K = 1200 \sim 2600$ is sufficient. We confirm that the spectral radius $\rho$ shows a complex influence on the prediction accuracy and suggest a smaller spectral radius $\rho$ currently. By changing the initial flow depth of dam break, we also obtained the conclusion that the prediction horizon of RC-ESN is larger than that of LSTM.

# 1. Introduction

With the incidence of recurring heavy rainfall, snowmelt, and other extreme abnormal weather around the world in recent years, the resulting floods have become increasingly widespread (Wu et al., 2014) . Among the flood hazards, dam-break floods have become a very important topic for engineers due to their sudden occurrence, rapid expansion, and urgent response. In the dynamics of dam-break floods, the wave propagation is responsible for catastrophic consequences such as the losses of life and properties in downstream areas (Schubert & Sanders, 2012) . Therefore, the computational prediction of wave propagation in dam-break floods is a long-standing problem in the engineering practice of hydrodynamics and hydrology.

The water wave propagation in dam-break floods is a dynamic process in which its spatial-temporal variation is usually more essential than the spatial pattern eventually formed (X. Li et al., 2017) . Conventionally, the wave propagation problem in dam-break floods can be solved by real experience and knowledge, reproduction of historical events and experiment tests, as well as physical and computational models (Aureli et al., 2021). A dam-break wave is a flow resulting from a sudden release of a mass of water in a channel. For most cases, wave propagation in a dam-break flood could be well described by the two-dimensional, depth-averaged Navier-Stokes (N-S) equations based on the shallow water assumption. For one-dimensional applications, the continuity and momentum conservation in the wave propagation process yields the Saint-Venant (S-V) equations (Barré de Saint-Venant, 1871), which are strongly related to the N-S equations.

Either the one-dimensional S-V equations or the two-dimensional simplified N-S equations, the governing laws describing wave propagation in the dam-break flood are expressed in the forms of hyperbolic partial differential equations (PDEs), requiring an adequate numerical method to solve these PDEs to obtain efficiency and accuracy results. Up-to-date, finite differential method (FDM), finite element method (FEM), and finite volume method (FVM) have found a large fan community (Seyedashraf et al., 2012) . Remarkable studies refer to, e.g., the Lax-Wendroff scheme (Lax & Wendroff, 1960), the leapfrog scheme (Fauzi & Memi Mayasari, 2021), and the scheme used in our previous study (Han et al., 2015) . Until now, the conventional PDEs-based numerical models are the dominant approaches for describing the wave propagation in dam-break floods, showing good agreement compared to the theoretical solution and flume experimental measurements. Nevertheless, in order to obtain results with high accuracy, these modeling techniques also demand extensive data sets. To solve this problem, some researches (Seyedashraf et al., 2018) are starting to try alternative approaches, rather than solve a problem directly.

Recently, artificial-intelligence-driven models based on the various deep neural networks have an increasing impact to assist research, for example, spatial and temporal forecasting of physical processes by solving PDEs. Due to its superior nonlinear approximation ability, deep learning neural networks have demonstrated obvious advantages for improving the nonlinear dynamics system simulation and prediction (Duraisamy et al., 2019; Gentine et al., 2018; Kutz, 2017; Peters, 2019; Reichstein et al., 2019; Scher & Messori, 2019; Schneider et al., 2017). One attraction of the machine-learning approach is respect to the acceleration and improvement ability of the prediction of complex dynamic systems (Chattopadhyay et al., 2020). Recently, deep learning neural networks are increasingly employed to assist in modeling chaotic (Pathak et al., 2018) and turbulent systems (Ling et al., 2016), and have yielded promising results (McDermott & Wikle, 2017, 2019; Raissi et al., 2019; Vlachas et al., 2018a). In this paper, deep learning method will also be used to make a preliminary attempt to one-dimensional dam-break wave propagation and directly predict the flow depth. Considering that this research is a preliminary exploration stage, the relatively simple one-dimensional dam break problem is studied here, and its reliable numerical solution can provide data for this research and reduce the influence of other factors.

Solutions of PDEs for wave propagation are in the form of time-series, spatio-temporal data, consisting of velocities and flow depths. In this sense, the long short-term memory (LSTM) model, which is most suitable for the prediction of sequence data, has attracted a lot of attention in recent years (Goodfellow I., 2016). The LSTM is the most prominent study in overcoming the difficulty that recurrent neural networks (RNNs) are difficult to learn long-term dependence on data and gradient disappearance and explosion problems (Bengio et al., 1993; Bynagari, 2020; Pascanu et al., 2013). For the topic of dam-break floods, a remarkable study refers to Fotiadis et al.(Fotiadis et al., 2020), who used LSTM as a benchmark for predicting surface wave propagation under two-dimensional images. In this paper, we also use LSTM to compare with the method in this paper. In contrast, we will directly use flow depth data for training instead of images.

This paper mainly uses the echo state network (ESN) (Jaeger, 2001) to predict the flow depth of one-dimension dam-break wave. The ESN, together with the liquid state machines (LSM) (Maass et al., 2002), is collectively the so-called reservoir computing (RC) (Verstraeten et al., 2007). Instead of a vast number of repetitions like in the back-propagation through time (BPTT) algorithm (Werbos, 1990), the RC-ESN employs a single training procedure. Notably, the RC-ESN's sophisticated network structure (reservoir) is connected by a large number of neurons, and the weights of the reservoir connection matrix need to be initialized in advance, enabling the RC-ESN better stability compared to the other neural networks as substantiated by many previous studies (D. Li et al., 2012; Lin et al., 2009; Skowronski & Harris, 2007; Tong et al., 2007).

Our goal is to predict the spatio-temporal propagation of water waves in the dam-break flood using a data-driven model based on the deep learning neural network. The reservoir computing echo state network (RC-ESN) is applied and trained with the sequence simulated results by solving the S-V equations for a one-dimensional dam-break flood process. To support the positive effect of the RC-ESN in long-term predicting, the LSTM model is selected as a comparison, predicting results by both two models will be compared to the numerical solution. The influence of related parameters on the prediction performance will be analyzed.

## 2. Materials and Methods

**2.1 PDEs and the numerical solutions for dam-break flood**

In this paper, we focus on the modeling of wave propagation in a one-dimensional dam-break flood scenario as shown in **Fig.1(a)**. This classic case has been employed as the benchmark test for many numerical studies (Han et al., 2015; Lhomme et al., 2010; Seyedashraf et al., 2012; Sheu & Fang, 2001), offering a good test benchmark for controlled analysis. To describe the wave propagation behavior, the Saint-Venant (S-V) equations are commonly used as the governing equations, owing that they exhibit a simplified mathematical structure without sacrificing the ability to consider smoothly flow conditions and flow discontinuities such as hydraulic jumps, moving bores, and the propagation on dry beds (Cozzolino et al., 2015). For the one-dimensional simplified dam break problem on the horizontal channel, the frictional force is ignored, and the expression is as follows:

$$\begin{cases} \dfrac{\partial}{\partial t}h + \dfrac{\partial}{\partial x}hv = 0 \\ \dfrac{\partial}{\partial t}hv + \dfrac{\partial}{\partial x}\left(hv^2 + \dfrac{gh^2}{2}\right) = 0 \end{cases} \tag{1}$$

where $h(x,t)$ is the flow depth at position $x$ at time $t$, $g = 1 kg/m^3$ is the gravity acceleration, and $v(x,t)$ is the propagation velocity along the $x$ direction.

In this one-dimensional case, as shown in **Fig.1(a)**, the dam-break flood is simulated in an assumed flume that is 20m in length. The bottom of the flume is horizontal, initially covered by a layer of 0.6m thick water in the downstream direction. On the upstream of the flume, a 1.8m-height dam exists, filling water on the upstream side. At the beginning of the simulation, the dam is suddenly removed, and the water in the reservoir is released onto the wet bed, generating a dam-break flood downstream.

The S-V equations belong to hyperbolic PDEs, and the exact theoretical solutions are available for some simple cases (Ancey et al., 2008; Thacker, 1981). Rather than the exact theoretical solution, we choose to train the deep neural network with a numerical solution because of its ability to adapt to more complex scenarios in future studies. Therefore, we solve **Eq. (1)** using the Lax-Wendroff method (Lax & Wendroff, 1960) in the MATLAB environment. In the solution, the grid number $n$ is set to 200, illustrating that the assumed 20m-long flume is separated into 200 grids with a uniform size of $\Delta x = 0.1m$. The reflective boundary condition on both sides of the flume is adopted. We solve the one-dimensional S-V equations with the following initial conditions, that is, the initial flow depth and velocity are

$$h_0 = \begin{cases} 1.8m, n \in [1,44] \\ 0.6m, n \in [45,200] \end{cases} \tag{2}$$

$$v_0 = 0 (n \in [1,200]) \tag{3}$$

To obtain a better simulation result, we set a relatively small-time increment $\Delta t = 0.001s$, that $100s$ of dam-break flood simulation consisting of 100,000 individual time steps. Simulation results containing $h(x,t)$ of each time step are well recorded (as illustrated in **Fig.1(b-c)**). All the simulated results have been uploaded to the GitHub repository for free access along with this paper.

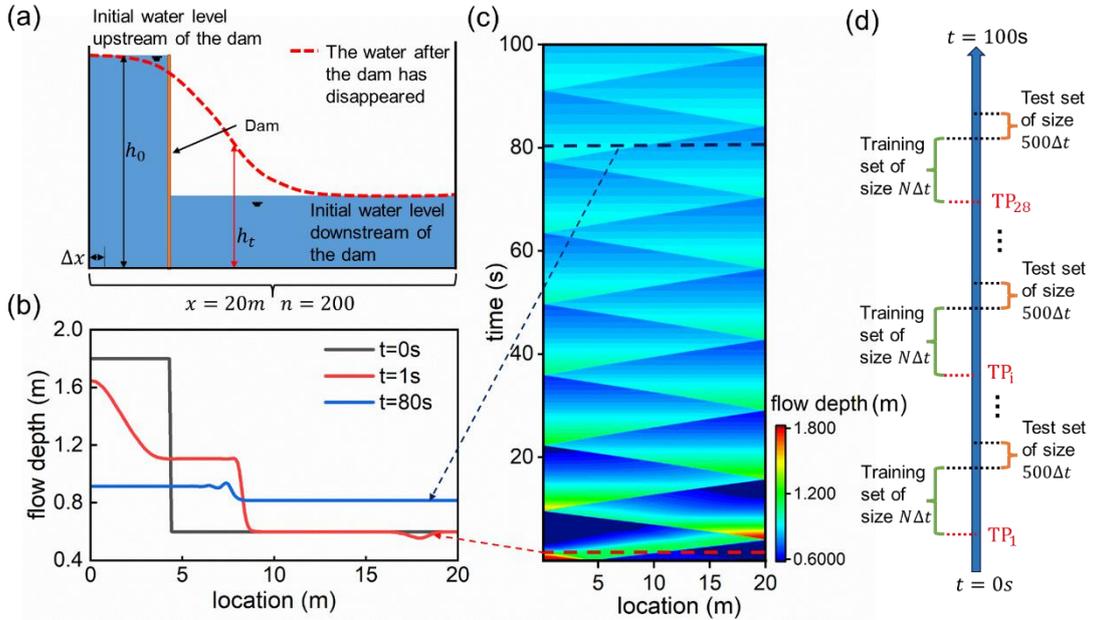

**Fig.1 Linear dam-break with an initially wet bed.** (a) the diagram of linear dam-break. (b) is the section shown by the dotted line in panel (c). (b) The flow depth at each position of the flood at the initial time $t = 0$, $t = 1s$, and $t = 80s$. (c) Time evolution of flow depth $h$. The abscissa is the position, the ordinate is the time, and the color bar is the flow depth. (d) The training and test datasets. Different datasets generate different time period (TP).

To generate the training and test datasets from the above numerical solutions, we sample the flow depth $X \in \Re^{200}$ at each time step $\Delta t$. The dimension of flow depth is 200, and the dimension of the input and output layers of the network is also 200. We construct a training set consisting of

$N = 2000$ sequence samples and a test set comprising the next 500 sequence samples from $N + 1$ to $N + 500$. In our study, we randomly selected 28 sets of such training/test data, each of them has a length of $(N + 500)\Delta t$. Different datasets generate different time period (TP) for training the machine learning model. Therefore, a total of 28 datasets, namely $\text{TP}_1$, $\text{TP}_2$, …, and $\text{TP}_{28}$, are tested, as shown in **Fig.1(d)**.

**2.2 The reservoir computing echo state network (RC-ESN) model**

As mentioned above, the RC-ESN (Jaeger, 2001; Jaeger & Haas, 2004) is a typical reservoir computing method. It consists of three layers, i.e., a randomly generated input layer, a high-dimensional sparse hidden layer, and an output layer that is unique to learn. Among them, the weights of the input layer and hidden layer are randomly sampled from a specified distribution and kept fixed in the training stage without learning, while the only weight of the output layer that requires learning may be solved simply by the regression method. As the core structure, the hidden layer is commonly referred to as a "reservoir". The structure diagram of the RC-ESN is shown in **Fig.2(a)**. In our study, we use the flow depth of each position at the last moment $X(t)$ as the input, so that the result of flow depth $X(t + \Delta t)$ at the next moment can be predicted using the RC-ESN model.

The reservoir has a size of $K$, and the connectivity of neurons in the reservoir is represented by the adjacency matrix $A$ with a size of $K \times K$, whose value is a random floating-point number on the interval [0,1). The adjacency matrix $A$ is normalized with its maximum eigenvalue $\Lambda_{max}$ and then multiplied by the spectral radius $\rho \leq 1$ to form the reservoir cyclic connection weight matrix $W_{res}$, which ensures that the reservoir system meets the necessary condition of stability, namely echo state property (Jaeger, 2001). The reservoir size $K = 1400$ and the spectral radius $\rho = 0.1$ are used in this study to obtain the best prediction performance. Sensitivity analysis of the prediction accuracy concerning $K$ and $\rho$ refers to the following discussion section.

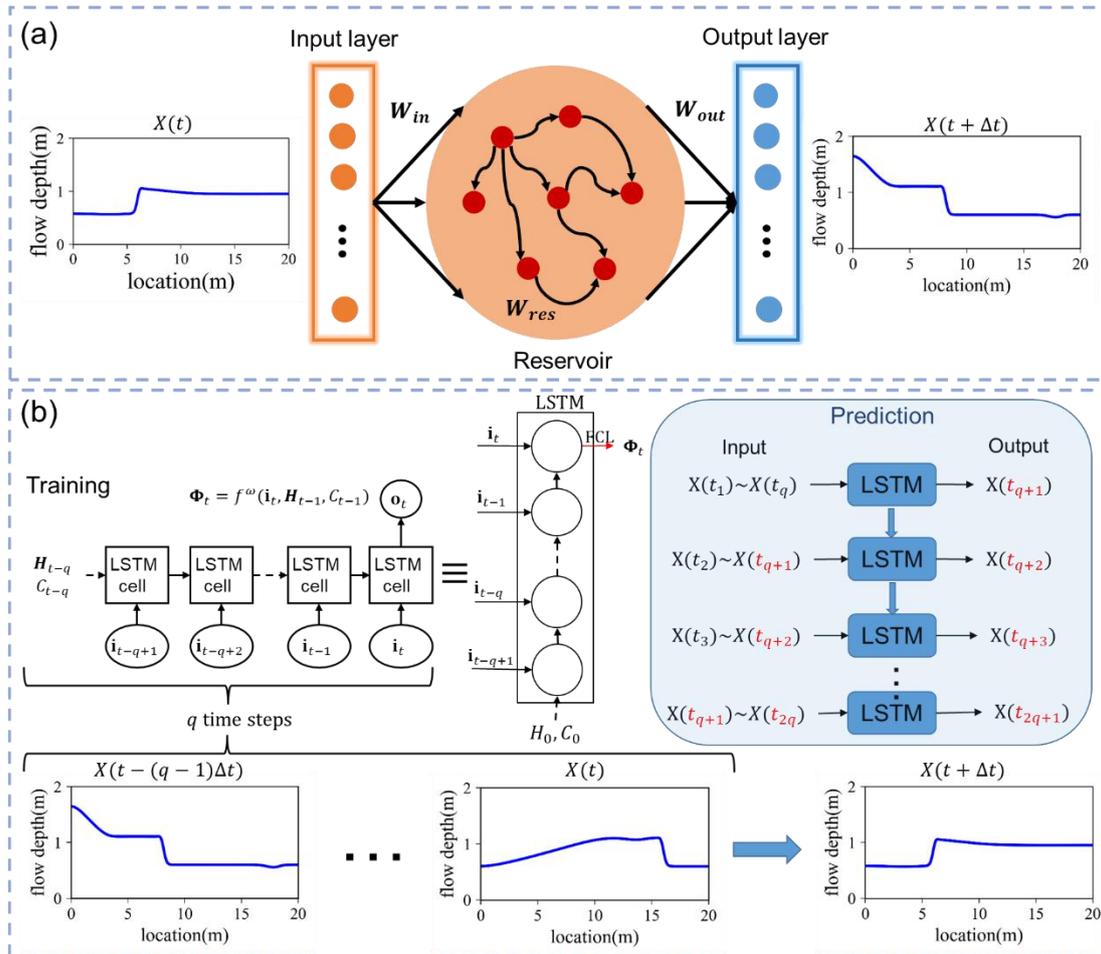

**Fig. 2 Schematic diagram of flow depth prediction by RC-ESN and LSTM methods.**
(a) Schematic diagram of ESN method. $X(t)$ and $X(t + \Delta t)$ are the input and output of ESN. During training, $X(t)$ is the known flow depth at the last moment and $X(t + \Delta t)$ is the known flow depth at the next moment. When predicting, $X(t)$ is the known flow depth or the predicted flow depth at the last moment. (b) An LSTM unfolded $q$ time steps and flow depth prediction. In the training stage, the flow depth of the previous $q$ time steps $[X(t-(q-1)\Delta t),...,X(t-\Delta t),X(t)]$ is taken as the input, and the output is the flow depth of the next moment $X(t+\Delta t)$. During prediction, the input can be known or previously predicted flow depth.

Input connection weight matrices $W_{in}$ and adjacency matrix $A$ are initialized with random numbers during training, which remains unchanged during training and testing, and $W_{res}$ is calculated. Only the weight matrix $W_{out}$ from the output layer to the reservoir is updated during training. The following is the governing equation of the echo state network throughout the training phase, where **Eq. (5)** is the echo state update formula and the echo state matrix $e$ is initialized to 0:

$$e_{t+\Delta t} = f(W_{res}e_t + W_{in}X(t)) \tag{5}$$

$$W_{out} = \arg\min_{W_{out}} \|W_{out}e_t - X(t)\| + \alpha\|W_{out}\| \tag{6}$$

where $X(t) \in \Re^{200}$ represents the input of time step $t$, which is either known initial conditions or has been predicted. The dimension of the input vector is $D$, where $D = 200$. The echo state generated by the echo state network at time step $t$ is represented by $e_t \in \Re^K$. $W_{res} \in \Re^{K \times K}, W_{in} \in \Re^{K \times D}$ and $W_{out} \in \Re^{D \times K}$ are the weight of the circular connection, input connection, and output connection, respectively. $f(\cdot)$ represents the activation function (e.g., $\tanh(\cdot)$) of the reservoir. $W_{out}$ can be solved simply by linear regression. Here, $\|\cdot\|$ is the vector's $L_2$-norm and $\alpha$ is the $L_2$ regularization constant.

For the prediction, the calculated $W_{out}$ is used to advance in time, and $e$ is updated continuously with the predicted $X$. The output of the prediction process is expressed as follows:

$$X(t + \Delta t) = W_{out} e_{t+\Delta t} \qquad (7)$$

## 2.3 The long short-term memory (LSTM) network

To further support the long-term prediction performance of the proposed RC-ESN model, the long short-term memory (LSTM) network, commonly used in many up-to-date studies (Alizadeh et al., 2021; Hayder et al., 2022), is selected as a comparison.

The input of LSTM is a column of the time-delay-embedded matrix $I_t$ of $X(t)$ with the embedding dimension $q$, commonly known as lookback (Kim et al., 1999) . In our study, the time-delay-embedded matrix has a dimension of $(200 \times q) \times M$, where $M$ is the column number of $N$ sequence samples $X \in \Re^{200}$ to form the delay matrix. The LSTM is calculated for forwarding propagation from a specific initial hidden state $H_0$ and an initial cell state $C_0$. Its cell state $C_t \in \Re^G$ and the hidden state $H_t \in \Re^G$ update formulas are presented below:

$$\widetilde{C}_t = \tanh(\mathbf{W}_C[\mathbf{H}_{t-1}, \mathbf{i}_t] + b_C) \qquad (8)$$

$$g_t^f = \sigma(\mathbf{W}_f[\mathbf{H}_{t-1}, \mathbf{i}_t] + b_f) \qquad (9)$$

$$g_t^i = \sigma(\mathbf{W}_i[\mathbf{H}_{t-1}, \mathbf{i}_t] + b_i) \qquad (10)$$

$$C_t = g_t^f C_{t-1} + g_t^i \widetilde{C}_t \qquad (11)$$

$$g_t^o = \sigma(\mathbf{W}_o[\mathbf{H}_{t-1}, \mathbf{i}_t] + b_o) \qquad (12)$$

$$\mathbf{H}_t = g_t^o \tanh(C_t) \qquad (13)$$

where $g_t^f, g_t^i, g_t^o \in \Re^{G\times(G+B)}$ are the gate signals (forget, input, and output gates), $\sigma$ represents the $Sigmoid(\cdot)$ activation function, which determines the weight ranging between 0 and 1. $\mathbf{i}_t \in \Re^B$ is the column of the time-delay-embedded matrix $I_t$ and represents the input of time step $t$. $B$ is the dimension of the input vector, which is $(200 \times q)$. $\mathbf{W}_C, \mathbf{W}_f, \mathbf{W}_i, \mathbf{W}_o \in \Re^{G\times(G+B)}$ and $b_C, b_f, b_i, b_o \in \Re^G$ represents the weight and bias respectively. The state's dimension $G$ is the number of hidden units, controlling the cell's ability to encode historical information (Vlachas et al., 2018b).

Assume that $\Phi_t$ is the output of LSTM, because the output in this study is supposed to have a specified dimension of $D = 200$, we add a fully connected layer (FCL) $\mathbf{W}_{oh} \in \Re^{D\times G}$ with no active function, as shown in **Eq. (14)**. We use the stateless LSTM, implying that the hidden and cell states of the LSTM are updated at the start of each batch of training. By expanding the LSTM of the previous $q$ time step and ignoring dependencies greater than $q$, **Eq. (15)** can be obtained,

$$\Phi_t = \mathbf{W}_{oh}\mathbf{H}_t = f^\omega(\mathbf{i}_t, \mathbf{H}_{t-1}, C_{t-1}) \tag{14}$$

$$X(t+\Delta t) \approx \Phi_t = \mathcal{F}^\omega(\underbrace{\mathbf{i}_t, \mathbf{i}_{t-1}, \ldots, \mathbf{i}_{t-q+1}}_{\mathbf{i}_{t:t-q+1}}, \mathbf{H}_{t-q}, C_{t-q}), \mathbf{h}_{t-q} = 0, C_{t-q} = 0 \tag{15}$$

where $\mathcal{F}^\omega$ represents the iterative application of $f^\omega$ and computation of the LSTM state for $q$ time steps, as shown in **Fig.2(b)**.

After debugging, the hidden layer number $l = 2$, hidden unit number $G = 800$, and lookback $q = 4$ are used in this study to obtain LSTM's prediction performance. LSTM predicts $X(t + \Delta t)$ based on the pre-$q$ time step of $X(t)$.

The weights $\mathbf{W}_C, \mathbf{W}_f, \mathbf{W}_i, \mathbf{W}_o, \mathbf{W}_{oh}$ of LSTM are learned using Adam (Kingma & Ba, 2015) optimizer via BPTT (Goodfellow I., 2016) algorithm during training. Here, the training samples in each batch are randomly shuffled to provide an unbiased gradient estimator in the stochastic gradient descent algorithm (Meng et al., 2019). During the prediction, $X(t + \Delta t)$ is predicted using $q$ past observations $[X(t - (q - 1)\Delta t),...,X(t - \Delta t),X(t)]$, which can be known or previously predicted. As shown in **Fig.2(b)**, $X(t_{q+1})$ is predicted by $X(t_1) \sim X(t_q)$, then, $X(t_2) \sim X(t_{q+1})$ predicts $X(t_{q+2})$, and so on.

### 2.4 The evaluation method for prediction performance

To ensure that optimal performance is not influenced by the initial conditions chosen, training/test sets with different initial conditions are sampled from the sample data independently and uniformly. We employ the root mean square error (RMSE) as a comparative measure. It is defined by **Eq. (16)** for each time period $TP$,

$$RMSE_{TP}(t) = \sqrt{1/n \sum_{i=1}^{n} \left(X_{TP}^i(t) - \widetilde{X}_{TP}^i(t)\right)^2} \qquad (16)$$

where $i$ represents the $i$th grid, and $n$ indicates the total number of grids. $X$ is the numerical solution data at time $t$, and $\widetilde{X}$ is the predicted value at time $t$ using one of the deep learning algorithms. RMSE was calculated for flow depth under different time periods at each moment, and an error curve describing the error's evolution through time was obtained.

In addition, the mean anomaly correlation coefficient (ACC) (Allgaier et al.,

2012) of 28 time periods is used to evaluate the pattern correlation between the predicted and numerical solution. ACC is defined as

$$ACC(t) = \frac{1}{28} \sum_{TP=1}^{28} \frac{\sum_{i=1}^{n}(X_{TP}^i(t) - \overline{X}_{TP}^i)(\widetilde{X}_{TP}^i(t) - \overline{X}_{TP}^i)}{\sqrt{\sum_{i=1}^{n}(X_{TP}^i(t) - \overline{X}_{TP}^i)^2 \sum_{i=1}^{n}(\widetilde{X}_{TP}^i(t) - \overline{X}_{TP}^i)^2}} \quad (17)$$

Here, the temporal average of flow depths at each grid is denoted by $\overline{X}_{TP}^i$. The score ranges from -1.0 to 1.0. if the prediction is accurate, the score equals 1.0.

## 3. Results

Considering the different frequencies of flow depth change in the development process of the dam-break flood, the prediction effect will be influenced by the flow depth at different moments. Therefore, as shown in **Fig.1(d)**, we randomly selected 28 training/test sets from different moments and set the size of the training set $N = 2000$. Sensitivity analysis of the prediction accuracy concerning $N$ refers to the following discussion section.

The prediction ability of RC-ESN and LSTM is compared in **Fig.3-5** using the identical training/test set. **Fig.3(a)** and **Fig.3(b)** illustrate the mean value of RMSE and ACC over time of RC-ESN and LSTM under 28 distinct randomly and uniformly selected time periods, respectively, to strengthen the comparison between the two methods and to avoid the influences by time periods. Results show that, with increasing prediction steps, the error of RC-ESN and LSTM increases progressively, and the model correlation between the predicted results and the simulation solution decreases continuously. However, in terms of prediction error and correlation between prediction results and numerical solution, RC-ESN outperforms LSTM in prediction performance.

RMSE increases gradually with the increase of the prediction time step. To display the prediction effect more intuitively, we call the prediction time step when RMSE reaches 0.01 as the prediction horizon, that is, within the prediction horizon, the error of the prediction effect is acceptable. **Fig.4** shows the RMSE values of the prediction and numerical solution as the prediction time step increase under two special time periods, i.e., the longest and shortest prediction horizon of RC-ESN and LSTM. **Fig.4(a)** shows that when the time period is $TP_1$, the prediction horizons of RC-ESN and LSTM method is the shortest, wherein, the prediction horizon of RC-ESN is 49, much better compared to the LSTM model; When the time period is $TP_{28}$, the RC-ESN method has the longest prediction horizon of 286, while the LSTM method has the longest prediction horizon of 81 when the time period is $TP_{26}$. In addition, **Fig.4(b)** compares RMSE for all time periods at the corresponding time for the above several prediction horizons. It can be seen that different time periods have a certain influence on the prediction effect, and the longer the prediction time step, the greater the influence.

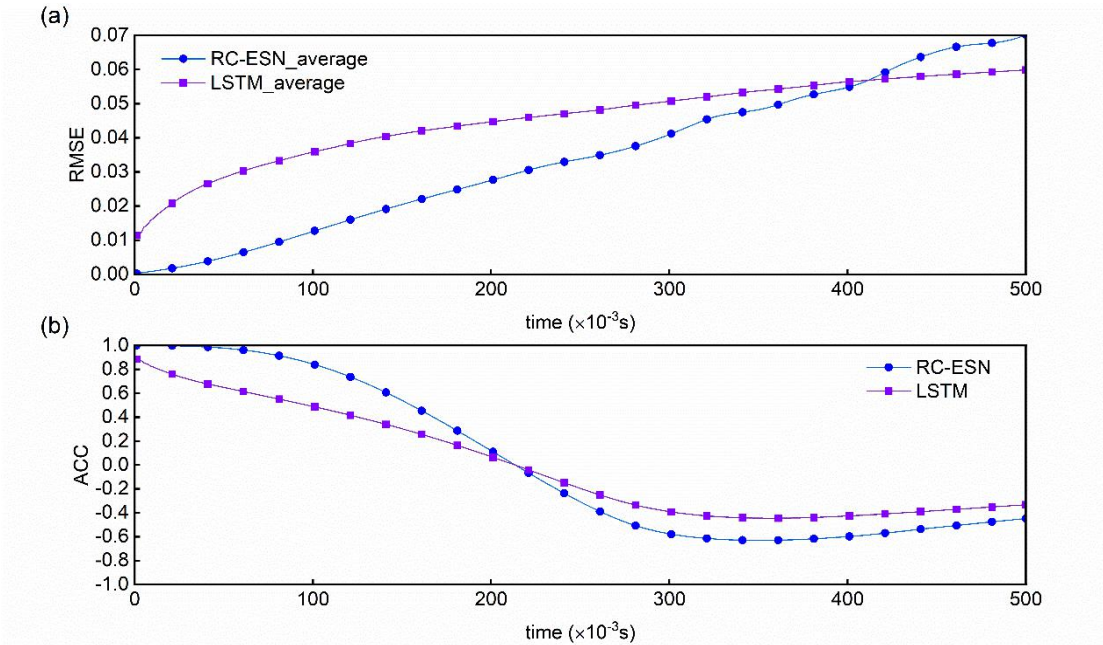

**Fig.3 Comparison of the prediction abilities among the two deep learning methods.** (a)The RMSE growth over time for RC-ESN (blue) and LSTM (purple) under all time periods. (b) The ACC growth over time for RC-ESN (blue) and LSTM (purple) under all time periods.

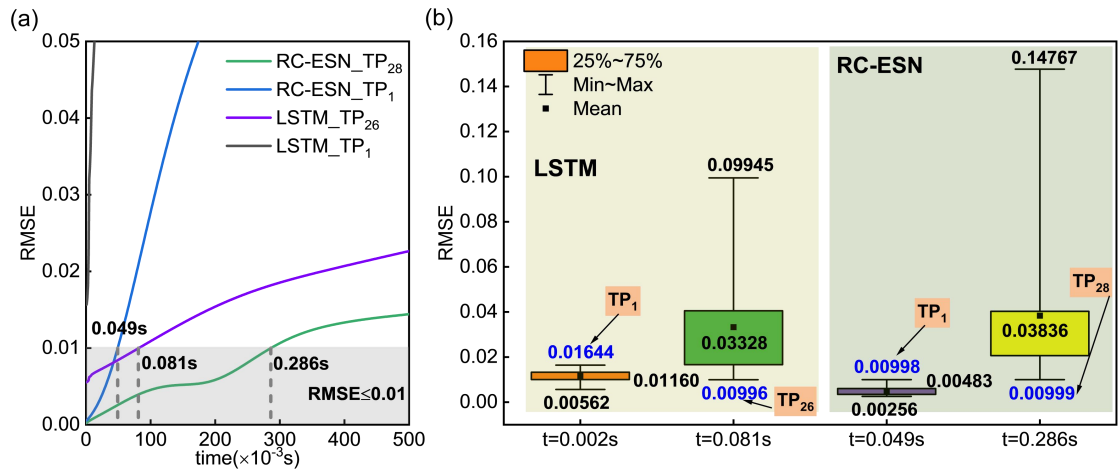

**Fig.4. The RMSE values of the prediction and numerical solution as the predicted time step increase under two special initial conditions.** (a) The change of RMSE with the prediction time step under the time period with the longest and shortest prediction horizon of RC-ESN and LSTM methods. (b) The RMSE at $t = 0.002s$, $t = 0.081s$, $t = 0.049s$, and $t = 0.286s$ under all time periods.

**Fig.5** presents the numerical solution and the prediction flow depth of the RC-ESN and LSTM method under the time periods and prediction horizons mentioned in **Fig.4**. Under the time period $TP_1$ that the prediction horizon of both models is the shortest, the figure shows the prediction effect of both models and the flow depth of dam-break waves at two instantaneous time steps, $t = 0.002s,$ and $t = 0.049s$. In this case, it is shown that although having the worst prediction effect under such time periods, the RC-ESN model can predict the wave propagation behavior because the predicted surface wave well reproduces the numerical solution. The figure also shows the prediction results under the time period $\text{TP}_{28}$, with which the RC-ESN attains the best performance, and the flow depth of numerical solution and both models at $t = 0.286s$. The figures demonstrate that the prediction result of RC-ESN is close to the numerical solution at $t = 0.286s$, whereas the prediction result of LSTM obviously deviates. Under the time period $\text{TP}_{26}$, with which the LSTM attains the best performance, although the local error fluctuation is observed in the RC-ESN's result at $t = 0.081s$, the overall RMSE of the RC-ESN model is lower owning to that the dam-break wave is predicted stagnant in the LSTM's results. It should be noticed that the error between the numerical solution and the prediction of both models gradually grows as the number of predicted steps increases.

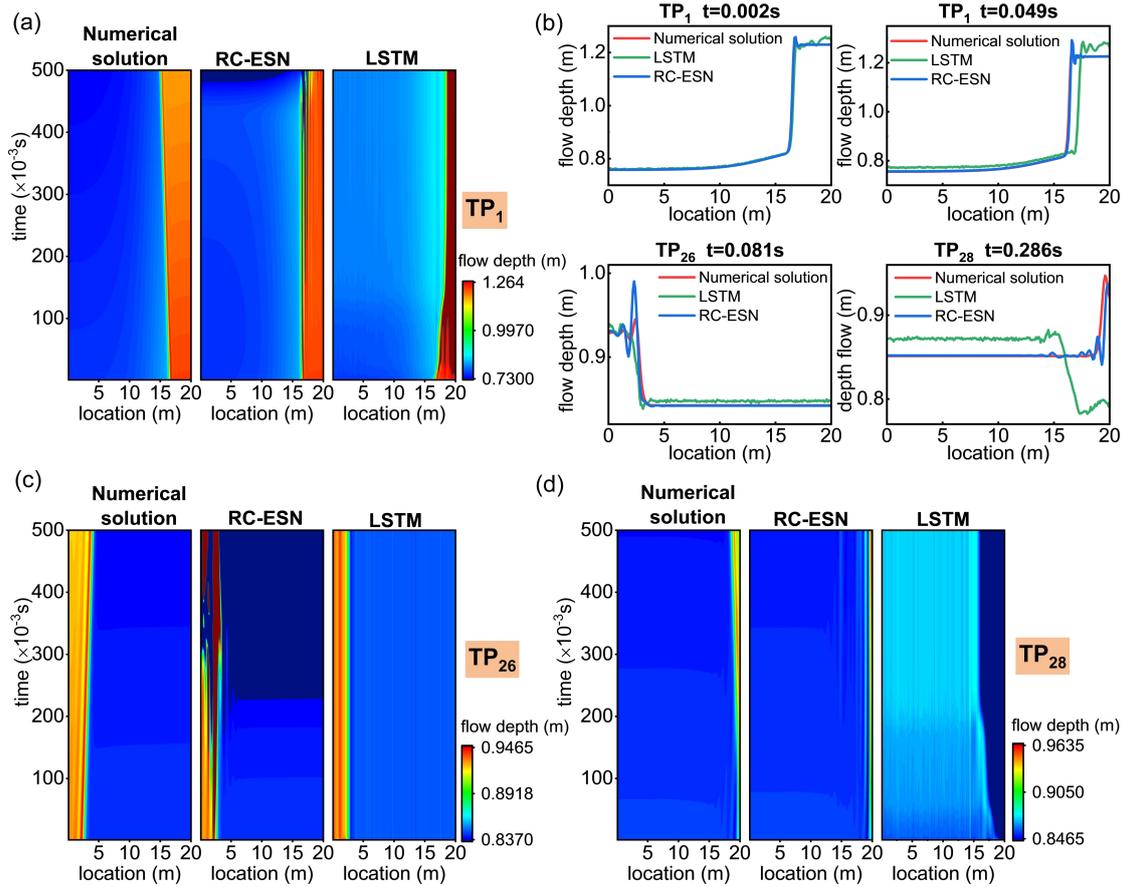

**Fig.5 Performance of RC-ESN and LSTM for spatiotemporal predictions.** The numerical solution, RC-ESN, and LSTM predicted flow depth under special time periods in **Fig.4** are shown above, respectively. (a), (c) and (d) are the flow depth at all locations and times under time periods $TP_1$, $TP_{28}$, and $TP_{26}$. The horizontal axis represents the location; the vertical axis represents the time (a total of $500\Delta t$ predicted, $\Delta t = 0.001s$); the color bar on the right represents the flow depth, in-unit $m$. (b) The numerical solution and predicted flow depth at each position at $t = 0.001s$, $t = 0.049s$, $t = 0.061s$, and $t = 0.286s$ under the time periods $TP_1$, $TP_{26}$ and $TP_{28}$.

As a conclusion of the result analysis, the RMSE of the RC-ESN reaches 0.01 later than that of LSTM regardless of the influence of the time periods, demonstrating that the RC-ESN model has a stronger ability for prediction compared to the conventional LSTM model.

## 4. Discussions

### 4.1. Sensitivity analysis on the training set size $N$

It is a long-standing issue in deep learning that how the model performance expands with the size of the training set has an important practical implication (Chattopadhyay et al., 2020) because the amount of data available for training is closely related to the prediction accuracy. The influence of both models, the RC-ESN and the LSTM, on the training set size from $N = 0.1 \times 10^4$ to $N = 8.3 \times 10^4$ is investigated. Here we extend the concept of prediction horizon as described above, as well as the prediction error $\overline{RMSE}$ is defined as the average of $RMSE$ between 0 and 100 prediction time steps ($0 \sim 100\Delta t$),

$$\overline{RMSE} = \frac{1}{100\Delta t} \sum_{i=0}^{i=100} RMSE(i\Delta t) \tag{18}$$

The prediction error $\overline{RMSE}$ and the prediction horizon are used to evaluate the abilities of both models for the prediction. **Fig.6(a)** shows how the $\overline{RMSE}$ varies as the size of the training set grows within 100 prediction time steps. It is shown that the expanded training data size is beneficial for improving the prediction ability because the prediction error $\overline{RMSE}$ of both models are reduced as the training size $N$ growing. In general, the $\overline{RMSE}$ of the RC-ESN model is lower than that of the LSTM method, implying that the RC-ESN model is less affected by the size of the training set.

**Fig.6(b)** demonstrates the variation of the prediction horizon against the training set size $N$. The prediction horizon of the RC-ESN and the LSTM both arise with the growth of training set size $N$. This positive effect of training set size $N$ to the prediction horizon is obvious for the RC-ESN model. The prediction horizon expands up towards 200 steps ahead when the training set size grows to $8.3 \times 10^4$. However, as to the LSTM model, this effect is not such obvious, with exists of several exceptional uncertainties when the training set size is beyond $4.5 \times 10^4$. In general, when compared to the LSTM, the prediction ability and the accuracy of RC-ESN are less dependent on the training set size, which is a significant advantage when the data set available for training is shorter, as in the case for predicting wave propagation in dam break floods.

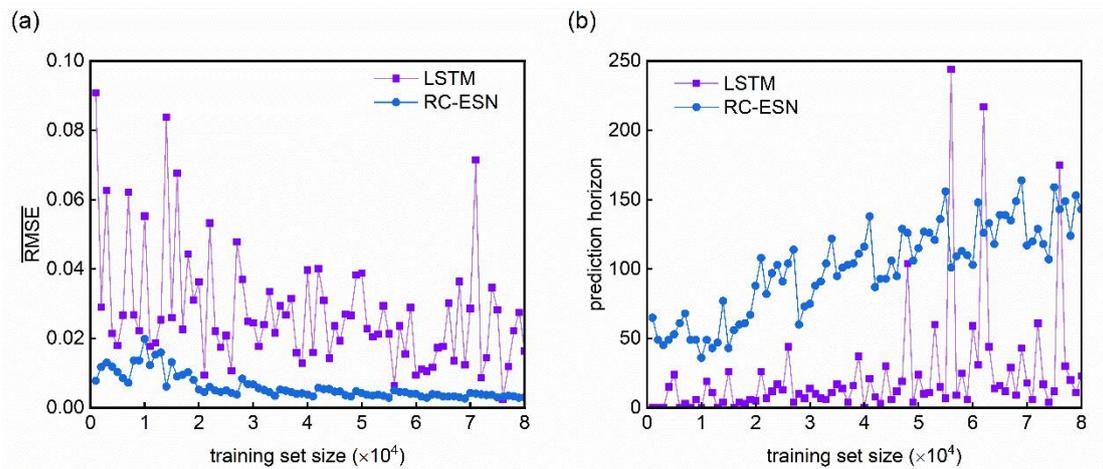

**Fig.6** Comparison of the prediction quality for RC-ESN (blue) and LSTM (purple) as the size of the training set $N$ is changed from $N = 0.1 \times 10^4$ to $N = 8.3 \times 10^4$. (a) The average error $\overline{RMSE}$ within 100 prediction time steps ($100\Delta t$). (b) The prediction horizon (when $RMSE = 0.01$).

**4.2. Sensitivity analysis on the reservoir size $K$ and spectral radius $\rho$**

The above analysis support that compared to the conventional LSTM model, the proposed RC-ESN model has witnessed a better performance for predicting wave propagation behavior of the one-dimensional dam-break flood. In this section, we focus on the other major concern of deep learning, i.e., the requirement for large reservoirs, which can be computationally taxing.

In fact, the low requirement for computational consumption has been demonstrated in many previous studies, e.g., as a potential disadvantage of ESNs versus LSTMs for training RNNs (Jaeger, 2007). In this study, we test the reservoir sizes $K$ ranging from 200 to 5000 to evaluate how the prediction error RMSE varies with the number of the prediction time. **Fig.7** depicts the prediction performance of the RC-ESN as reservoir size $K$ varies. **Fig.7(a)** shows the prediction horizon of the RC-ESN using the same initial conditions but different reservoir sizes. **Fig.7(b)** shows the variation of the RMSE with predicted time steps, and **Fig.7(c)** is an expanded view of the shaded region in **Fig.7(b)**. It is obvious that the number of predicted time steps gradually increases with the reservoir size, demonstrating an improvement of the RC-ESN's prediction performance. However, when the reservoir size $K$ grows beyond 2600 substantially, its positive effect for prediction gets weakened, as well as the computational efficiency decreases. In this sense, we conclude that although an increasing reservoir size $K$ is theoretically favorable for improving prediction performance, a medium reservoir size $K = 1200 \sim 2600$ is sufficient, considering the overall balance of computational efficiency and prediction performance.

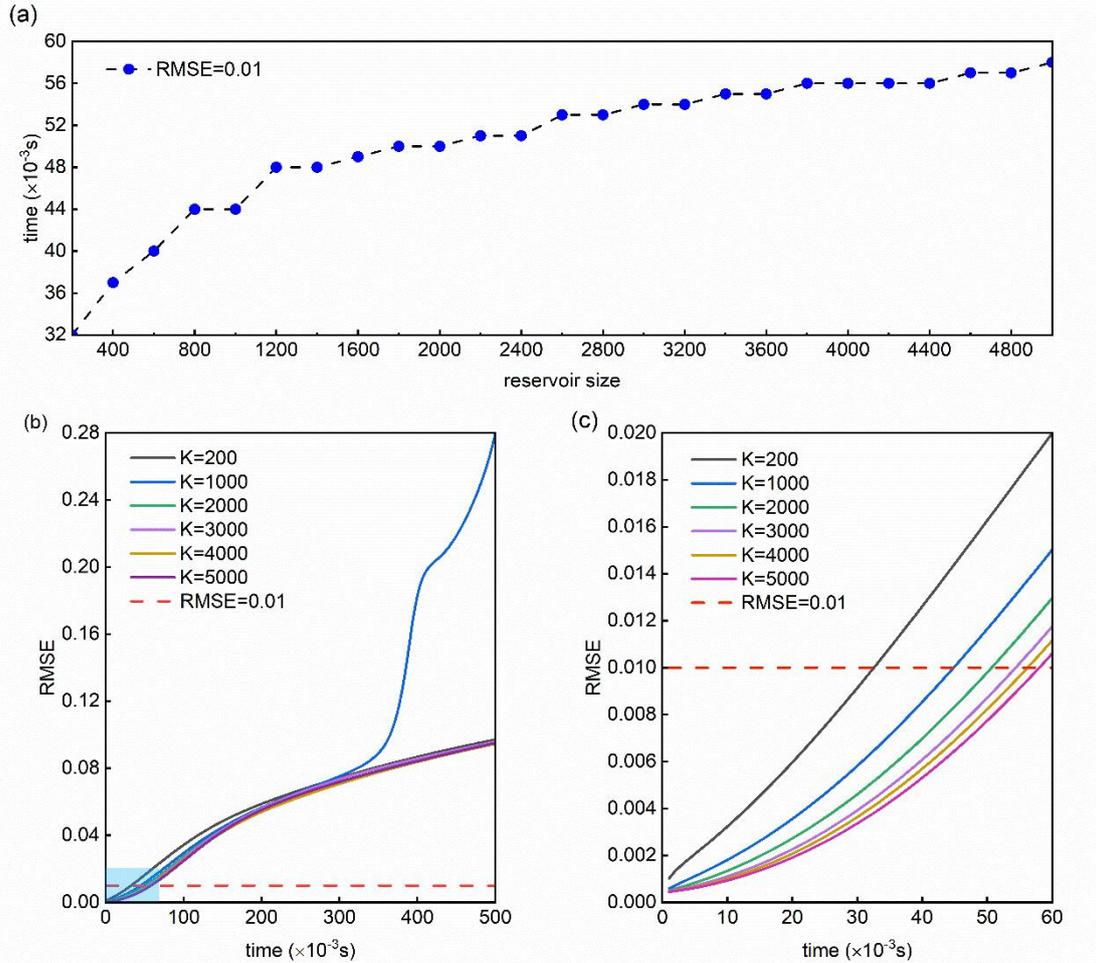

**Fig.7 Scaling of RC-ESN performance with reservoir size $K$.** (a) Under the same initial conditions, the prediction horizon when the reservoir size changes from 200 to 5000. (b) When $K = 200\sim5000$, the RMSE of the prediction results under the same initial conditions changes with the prediction time. Panel (c) is the enlargement of the shaded part in the lower-left corner of the panel (b), where the red dotted line is $RMSE = 0.01$.

Besides the reservoir size $K$, we also explore the influence of the spectral radius $\rho$. Some studies uncover the emergence of an interval ("valley") in the spectral radius of the neural network in which the prediction error is minimized, and such an interval arises for a variety of spatiotemporal dynamical systems described by nonlinear partial differential equations, regardless of the structure and edge-weight distribution of the underlying reservoir network (Jiang & Lai, 2019). In the following analysis, the best-fitting reservoir size $K = 1400$ is used. We test different spectral radius $\rho$ ranging from 0.01 to 1.00 under the identical time period $TP_{28}$, then compared its impacts to the prediction results. **Fig.8(a)** shows the RMSE variation in different prediction time steps. It is found that the influence of spectral radius is limited for the cases of predicted time steps less than 200. However, as shown in the figure, the curves begin to disperse from each other once the predicted time step is beyond 200 steps, demonstrating that significant differences in RMES appear. As the curves in **Fig.8(a)** are mixed, to have a better exploration of the influence of the spectral radius $\rho$, we separately display the curves of $\rho = 0.01{\sim}0.3$, $\rho = 0.31{\sim}0.66$ and $\rho = 0.67{\sim}1.0$ in the panels of **Fig.8(b-d),** respectively. As illustrated in **Fig.8(b)** and **Fig.8(c)**, when $\rho = 0.01{\sim}0.66$, the RMSE value of the predicted results increases. However, when spectral radius $\rho$ continually grows, $\rho = 0.67{\sim}1.0$, a reverse trend is observed, that the RMSE diminishes progressively as the spectral radius increases as demonstrated in **Fig.8(d)**. This phenomenon implies that the spectral radius $\rho$ has a complex influence on the prediction performance of the RC-ESN model, which deserves ongoing studies, in the future in particular for the tasks of wave propagation prediction in dam-break floods. In general, considering that $\rho = 0.01$ obtains the best result to control the accumulation of RMES in **Fig.8(a)**, we currently suggest using a smaller spectral radius $\rho$ for a better prediction performance.

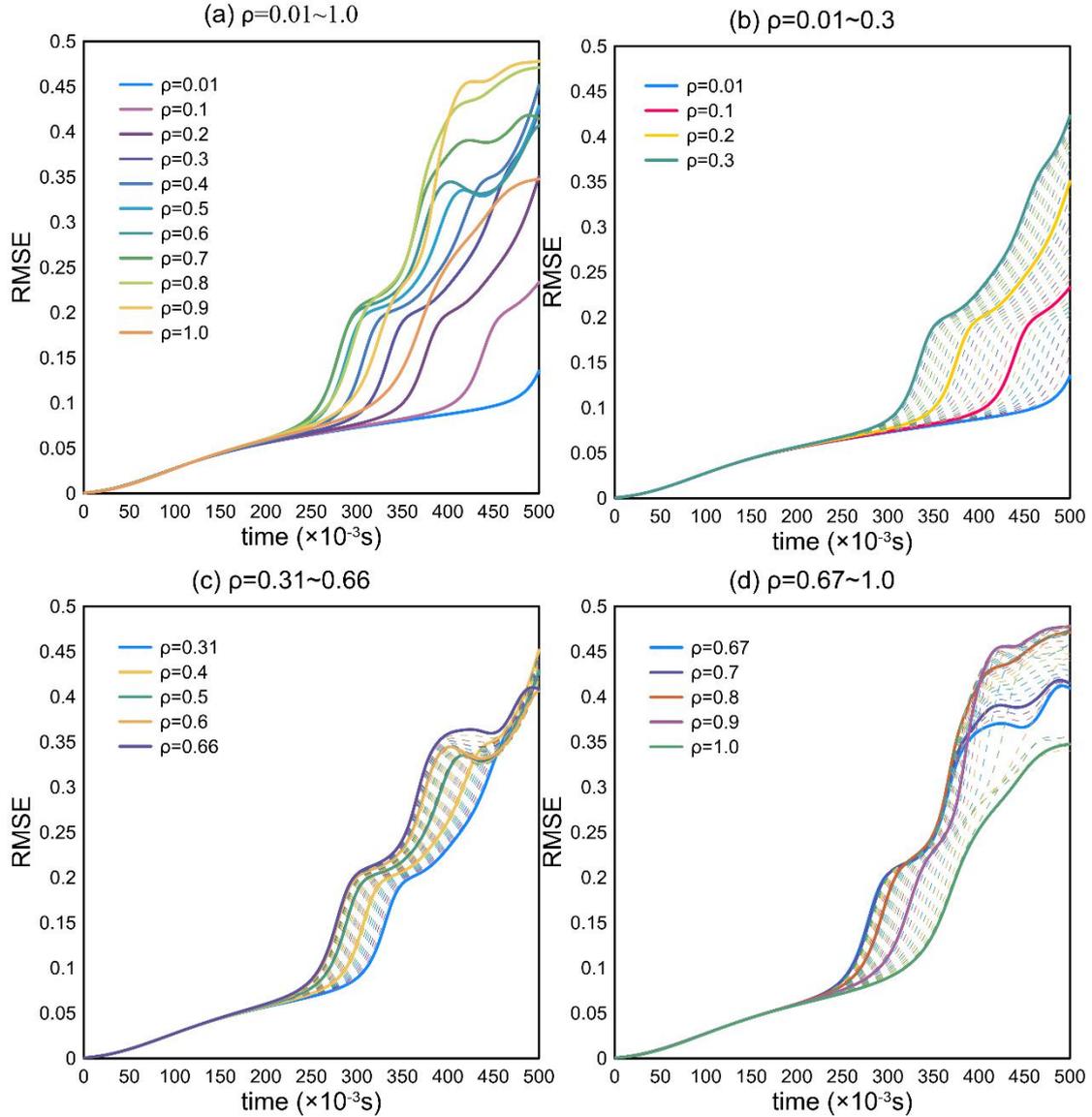

**Fig.8 Scaling of RC-ESN performance with spectral radius $\rho$.** The spacing of dotted lines is $\rho = 0.01$.

**4.3. Dam-break wave prediction under different initial flow depths**

The method proposed in this paper is used to train and predict the propagation of dam-break waves with initial flow depths $h_0^1$ and $h_0^2$, respectively. Other parameters are consistent with those described above.

$$h_0^1 = \begin{cases} 1.5m, n \in [1,28] \\ 0.8m, n \in [29,200] \end{cases} \quad (19)$$

$$h_0^2 = \begin{cases} 1.7m, n \in [51,70] \\ 0.6m, n \in [1,50] \cup [71,200] \end{cases} \tag{20}$$

**Fig.9(a)** shows the prediction horizons of RC-ESN and LSTM at different time period datasets under initial flow depths $h_0^1$ and $h_0^2$, respectively. **Fig.9(b)** and **Fig.9(c)** respectively show the predicted flow depths of LSTM at the time of maximum prediction horizon under $h_0^1$ and $h_0^2$ ($t = 0.052s$ and $t = 0.104s$), as well as the numerical solution and RC-ESN prediction results at this time. **Fig.9(d)** and **Fig.(e)** show the predicted flow depths of RC-ESN at the time of the maximum prediction horizon under $h_0^1$ and $h_0^2$ ($t = 0.162s$ and $t = 0.211s$), as well as the numerical solution and RC-ESN prediction results at this time. As can be seen from **Fig.9**, after the initial flow depth changes, the prediction horizon of RC-ESN in different time period datasets is larger than that of LSTM.

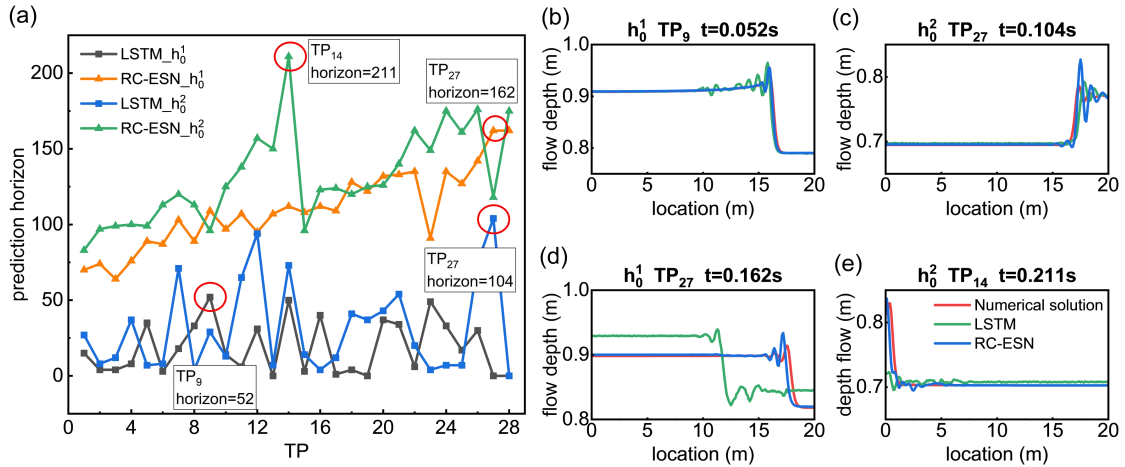

**Fig.9 Performance of RC-ESN and LSTM under different initial flow depths.** (a) The prediction horizons of LSTM and RC-ESN on 28 different time period datasets of the initial flow depths $h_0^1$ and $h_0^2$, respectively. (b) The flow depth at the time of maximum prediction horizon of LSTM under initial flow depths $h_0^1$. (c) The flow depth at the time of maximum prediction horizon of LSTM under initial flow depths $h_0^2$. (d) The flow depth at the time of maximum prediction horizon of RC-ESN under initial flow depths $h_0^1$. (e) The flow depth at the time of maximum prediction horizon of RC-ESN under initial flow depths $h_0^2$.

## 5. Conclusions

In this study, we present that the data-driven RC-ESN model, that well-trained using the numerical solutions of Saint-Venant equations, can help predict the long-term dynamic behavior of a one-dimensional dam-break flood with satisfactory accuracy. Comparison with the numerical solution demonstrates that the proposed RC-ESN model achieves the best performance of ahead predicting wave propagation 286 time steps among the all 28 time periods, with a root mean square error (RMSE) smaller than 0.01. It is outperforming the conventional LSTM model which reaches a comparable RMSE only 81 time-steps ahead.

In the sensitivity analysis, we support that compared to the conventional LSTM model, the proposed RC-ESN model is less dependent on the training set size, which is a significant advantage when the dataset is available for training. For the purpose of wave propagation prediction, we find that although an increasing reservoir size $K$ is theoretically positive for improving prediction performance, a medium reservoir size $K = 1200 \sim 2600$ is sufficient, considering the overall balance of computational efficiency and prediction performance. However, it is confirmed that the spectral radius $\rho$ has a bit complex influence on the prediction performance of the RC-ESN model. We suggest using $\rho = 0.01$ or a smaller spectral radius $\rho$ for a better prediction performance in this one-dimensional dam-break flood scenario. In order to verify the above conclusion, we change the initial flow depth of the dam break problem, and still obtain the prediction horizon of RC-ESN is larger than that of LSTM in different time periods.

Our findings support a favorable role of the data-driven and machine-learning model to predict the wave propagation in a one-dimensional dam-break flood. However, for a two-dimensional or even three-dimensional scenario, which is closer to the real engineering problem, more training data, such as the wave propagation velocities along with different directions, should be involved and therefore, arising the difficulties for training a well-performance neural network. This aspect obviously needs more work and should be continued investigated.

## Data availability

The data that support the findings of this study are available in the GitHub repository [lcl1527/dam-break-ESN-LSTM], at [lcl1527/dam-break-ESN-LSTM (github.com)](github.com). The numerical simulation is performed using MATLAB2019. Rc-ESN and LSTM training and prediction are performed using Pytorch. The source codes used in this work are freely available online in the Github repository: [lcl1527/dam-break-ESN-LSTM (github.com)](github.com).

## Acknowledgments

This study was financially supported by the National Key R&D Program of

China (Grant No. 2018YFD1100401); the National Natural Science Foundation of China (Grant No. 52078493); the Natural Science Foundation for Excellent Young Scholars of Hunan (Grant No. 2021JJ20057); the Innovation Provincial Program of Hunan Province (Grant No. 2020RC3002); and the Fundamental Research Funds for the Central Universities of Central South University (Grant No.2022ZZTS0660). These financial supports are gratefully acknowledged.